# $Fe_{5-x}Ge_2Te_2$ - A new exfoliable itinerant ferromagnet with high Curie temperature and large perpendicular magnetic anisotropy

Palani R. Jothi,[a] Jan P. Scheifers,[a] Yuemei Zhang,[a,c] Mohammed Alghamdi,[b] Dejan Stekovic,[a]

Mikhail Itkis,[a,d] Jing Shi,[b] Boniface P.T. Fokwa[a*]

[a] Department of Chemistry, University of California, Riverside, California 92521, USA

[b] Department of Physics and Astronomy, University of California, Riverside, California 92521, USA

[c] Department of Chemistry, Warren Wilson College, North Caroline 28778, USA

[d] Department of Chemical and Environmental Engineering, University of California, Riverside, CA 92521, United States

*Corresponding Author - E-mail: bfokwa@ucr.edu

**Abstract:**

Layered van der Waals (vdW) crystals with intrinsic magnetic properties such as high Curie temperature ($T_C$) and large perpendicular magnetic anisotropy (PMA) are key to the development and application of spintronic devices. The ferromagnetic vdW metal $Fe_{3-x}GeTe_2$ (FGT) has gained prominence recently due to its high $T_C$ (220 K) and strong PMA. Here, we introduce a new metallic vdW ferromagnets, $Fe_{5-x}Ge_2Te_2$ or FG2T, which was successfully synthesized and fully characterized. FG2T is a metal that orders ferromagnetically with a very sharp transition at 250 K (bulk and single crystal thin flakes) and shows large PMA, as found by both experimental and computational studies. This work enables novel heterostructure devices with near room temperature capabilities by using FG2T as spin injector.

**KEYWORDS**: van der Waals materials; exfoliation; ferromagnetism; PMA; DFT calculations

Layered van der Waals (vdW) materials have attracted a great deal of interest since the discovery of graphene through exfoliation of graphite. [1,2] The prospect of realizing nanoscale spintronic devices constructed from atomically thin two-dimensional (2D) ferromagnetic (FM) materials through combination of large spin polarization and high Curie temperature ($T_C$) has propelled these materials to a high level of interest in the scientific community. The first intrinsically 2D FM materials were realized from exfoliation of the ferromagnetic (FM) $CrI_3$ (insulator with bulk $T_C$ = 61 K) [3], FM $Cr_2Ge_2Te_6$ (insulator with bulk $T_C$ = 60 K) [4] and recently FM $VI_3$ (semiconductor with bulk $T_C$ = 50 K). [5] However, $T_C$ of these materials remain somewhat low for spintronic applications, thus creating new challenges for both experimentalists and theoreticians to search for new 2D intrinsic ferromagnets with much higher $T_C$. In contrast, itinerant FM vdW $Fe_{3-x}GeTe_2$ (FGT), discovered in 2006 by Deiseroth et al.,[6] has a high $T_C$ ranging from 150 K to 220 K (depending on Fe vacancy x) [6-10] and thus has recently attracted the attention of the scientific community. While bulk FGT is already very interesting as an itinerant FM showing Kondo lattice behavior, large anomalous Hall current among other properties, [6-9] it is the high $T_C$ and the large perpendicular magnetic anisotropy (PMA) in thin films that have propelled this phase to prominence. [10] Even more impressive is the recent discovery of 2D itinerant ferromagnetism in the FGT monolayer with the highest reported $T_C$ (130 K), [11] thus enabling the engineering of advanced spintronic vdW heterostructures. Some heterostructures have recently been realized by combining FGT with other vdW materials (h-BN, graphite) or metal thin film (Pt) to investigate various properties such as tunneling spin valves, antisymmetric magnetoresistance and spin-orbit torques (SOT). [12] While FGT has comparably high $T_C$, it is still below room temperature, however it was recently demonstrated that an ionic gate (using $Li^+$ intercalation) or microstructures patterning (by a focused ion beam) can raise $T_C$ of FGT thin flakes up to room temperature, thus creating further opportunities for room-temperature spintronics devices based on atomically thin van der Waals crystals. Nevertheless, it remains a challenge to find vdW crystals with high enough $T_C$ that will enable the exfoliation of monolayer with intrinsic above room temperature $T_C$, a prerequisite for any application. Johrendt et al. [14a] and McGuire et al. [14b] have recently found that another known vdW crystal in the Fe-Ge-Te system, $Fe_{5-\delta}GeTe_2$, [15] has an even higher $T_C$ than FGT, with values of 290 K and 310 K reported for the bulk crystals that show a high level of stacking fault disorder if compared to FGT crystals. $Fe_{5-\delta}GeTe_2$ structurally differs from $Fe_{3-x}GeTe_2$ because it has an extra honeycomb iron layer that contributes to the additional iron in the chemical formula.

We report on a new vdW crystal in the Fe-Ge-Te system, $Fe_{5-\delta}Ge_2Te_2$ (FG2T), that has a higher $T_C$ than FGT with a sharp FM transition at 250 K. Hall measurements on thin flake single crystals demonstrate large PMA and resistivity measurements show metallic behavior. Density functional theory (DFT) calculations have confirmed the itinerant FM ground state as well as strong PMA for thin flakes.

**Synthesis and crystal structure of $Fe_{4.84(1)}Ge_2Te_2$ (FG2T)**: FG2T was synthesized following a solid state reaction route from its elemental constituents. Two samples were submitted to different heat treatments: The first sample was annealed at 800 °C followed by slow cooling which yielded a polycrystalline powder containing small crystals, while the second sample was quenched after annealing at 760 °C leading to larger crystals. In both synthesis conditions the obtained products contained FG2T as majority phase as exemplified by the Rietveld refinement conducted on the powder X-ray diffraction (PXRD) data (see **Figure S1, Table S1** and the synthesis section in the SI). The crystal structure of FG2T was determined by single-crystal X-ray diffraction (SCXRD) from which the trigonal centrosymmetric space group $P\bar{3}m1$ (no. 164) was identified. The lattice parameters [$a = b$ = 4.0121(3) Å and $c$ = 10.7777(8) Å] obtained from the SCXRD analysis were in very good agreement with those from the two PXRD datasets (see **Table S2**) indicating a uniform sample as confirmed by the refined chemical formulas in all cases. The refined structure has 5 main Wyckoff positions in its asymmetric unit [Te (site 2d), Ge (site *2d*), Fe1 (site *2d*), Fe2 (site *2c*) and Fe3 (site *1a*)] leading to the initial chemical formula $Fe_5Ge_2Te_2$. However, a significant electron density was found at site *1b*, and a free refinement of this site's occupancy with Fe (like in the case of Ni in $Ni_{3-x}GeTe_2$ [5]) led to a partially occupied site (Fe4) with an occupancy of 16.4(9) %. Also, in all related phases found until now, the *2d* site (Fe1) is always reported as partially occupied, thus we have refined this site leading to an occupancy of only 83.8(7)%. Furthermore, we have observed enlarged (in the *ab* plane) anisotropic displacement parameters (ADPs) for Ge. Additionally, small peaks in the electron density map were observed at ca. 0.6 Å from Ge (also found in our electron localizability function (ELF) analysis, **cf. Figure S2**). Introducing a split site for Ge (73% at *2d* and 8.9% at *6i*) resulted in improved *R*-values, decreased ADPs and disappearance of the peaks in the electron density. The final refinement (see **Tables S3-S5**) led to excellent reliability values and a final composition of $Fe_{4.84(1)}Ge_2Te_2$, which is in perfect agreement with the refined compositions obtained from the Rietveld refinements of the two bulk samples (**Table S2**). The presence of the three elements was further confirmed by semiquantitative energy dispersive X-ray spectroscopy (EDS) done on several crystals **(Figure S3)**. In contrast to $Fe_{5-x}GeTe_2$ [14], we did not observe severe stacking disorder along [001] in any of the analyzed single crystals of FG2T (**Figure S4**). The final crystal structure of FG2T is presented on **Figure 1**. The Wyckoff sequence *164,id²cba* indicates a new structure type. The new structure is built from slabs stacked along [001] and separated by a vdW gap of 2.886(2) Å, a value slightly smaller than those of FGT (2.95 Å) [6] and $Fe_{5-x}Ge_2Te_2$ (3.06 Å).[17] Each FG2T slab consists of seven layers in the sequence ABCB'C'BA': Three layers of Fe (Fe2 in each of the two B and Fe3 in B') are stacked on top of each other, leading to face-sharing trigonal prisms of Fe2 and Fe3 atoms. These prisms are alternatively filled by Fe1 and Ge, resulting in the mixed Ge/Fe layers C and C'. The slab is terminated by a layer of Te on each side, namely A and A' layers. Due to the double layer of Fe-prisms and the alternate arrangement of Ge and Fe1 atoms within the double layer, the layers A and A' of the same slab are offset to each other. This has two important consequences: (1) only a single slab is required per unit cell to achieve translational symmetry along [001], and (2) the Te atoms of two neighboring slabs form empty octahedra and tetrahedra at the vdW gap. In the center of the octahedra, we found significant electron density reflecting a partially occupied (16% Fe) atomic position at (0,0,0). We denote this layer D.

Examining the structure of FG2T ($Fe_{5-x}Ge_2Te_2$) more closely reveals that the layers C and C' are not perfectly flat; that is Fe1 and Ge do not have the same *z*-coordinate. In fact, the Fe1-Ge distance is rather short (2.325 Å in average), which is the reason for the vacancies on the Fe1 site, the puckering of the layer and the split site for Ge (in the *ab* plane). This contrasts with $Fe_{5-x}GeTe_2$, [14] where Ge is displaced along *c* instead of in the *ab* plane. The displacement of the Fe1 atom towards the vdW gap increases the Fe1-Ge distance and shortens the Fe1-Te distance, which is even shorter than the shortest Fe2-Te distance of 2.563(3) Å. The short Fe1-Te distance reflects a strong bond, which explains why Te prefers to cap the Fe-filled trigonal prisms over the Ge-filled ones.

The FG2T structure is closely related to other layered tellurides and especially to FGT ($Fe_{3-x}GeTe_2$). The main difference between FGT and FG2T is the number of Ge/Fe layers building a slab: While FGT contains a thin slab of 5 layers centered around a single Ge/Fe layer C in a layer sequence ABCBA, FG2T exhibits a thicker 7 layers slab containing two Ge/Fe layers in a stacking sequence ABCB'C'BA'. Removing one Fe layer (B') and one $Fe_{1-x}$Ge layer (C'), i.e. one $Fe_{2-x}$Ge block, from $Fe_{5-x}Ge_2Te_2$ would lead to the FGT composition $Fe_{3-x}GeTe_2$.

**Density functional theory (DFT) results for FG2T**: Single-crystal X-ray diffraction of FG2T found partially occupied sites for Fe1 (84%) and Fe4 (16%) as well as a split site for Ge, all of which are not easy to simulate computationally. Therefore, a simplified model for $Fe_5Ge_2Te_2$, containing fully occupied Fe1, no Fe4 and no Ge split site (see **Table S5**), was used for our DFT total energy calculations (see SI for details). [21, 22] Density of states (DOS) from LDA calculations (**Figure 2 left**) shows a large density at the Fermi level ($E_F$) indicating electronic instability. By applying spin polarization (LSDA calculations), the split of majority and minority spins opens a pseudogap at $E_F$ (**Figure 2 left**), diminishes the electronic instability and induces magnetic ordering. Moreover, there is no band gap at $E_F$ for both LDA and LSDA calculations, indicating a metallic behavior for $Fe_5Ge_2Te_2$, as confirmed by the transport measurements (**Figure 2 bottom right**). To explore the partial occupancy of Fe1 site, chemical bonding analysis, using the crystal orbital Hamilton population (COHP) [23] curves (**Figure 1, right**), was applied for Fe1-Fe2, Fe1-Fe4, Fe1-Te and Fe1-Ge interactions. Except for the Fe1-Ge bond, all others have partially filled antibonding states. Fe1 partial occupancy can decrease the number of these bonds because it reduces the valence electron count, which moves $E_F$ to the left and reduces the filling of the antibonding region, thereby stabilizing the electronic structure. Consequently, partial occupancy of Fe1 would be expected. In addition, $E_F$ sits on a peak in the LDA-DOS and in the antibonding region of Fe1-Fe2 and Fe1-Fe4 COHP curves. Therefore, ferromagnetic (FM) couplings are expected for Fe1-Fe2 and Fe1-Fe4 interactions according to the COHP interpretation of magnetic interactions, [24] suggesting FM ordering for this compound. Furthermore, we have calculated the total energy of two magnetic models, one with FM couplings within the slab and between slabs, another containing antiferromagnetically coupled FM slabs. Our results show that the first model is more stable than the second by 13.6 meV/f.u. Therefore, the inter-slab magnetic coupling in $Fe_5Ge_2Te_2$ is weak FM unlike in $Fe_3GeTe_2$ where AFM is more stable (our own calculations and others [9c]), indicating FM ordering in this new compound. Indeed, magnetic measurements have confirmed this finding (see below and **Figures 2 and 3**). Lastly, the magnetic anisotropy of bulk and thin film FG2T was studied using spin orbit coupling (SOC) calculations. The magnetic anisotropy energy

(MAE = $E_{SOC}(\|c) - E_{SOC}(\perp c)$) obtained for the bulk is -3.38 meV/f.u. (-3.6 MJ/m$^3$), indicating very strong easy axis anisotropy. MAE as function of slab thickness was also examined for thin films. The results are plotted in **Figure 3 (left)**. The single layer has more than 10 times weaker anisotropy compared with bulk Fe$_5$Ge$_2$Te$_2$. From a single layer to a two-layers film, the anisotropy increases dramatically and then slightly increase as the number of layer increases. Therefore, it is predicted that any Fe$_5$Ge$_2$Te$_2$ thin film with more than one slab would have large easy axis magnetic anisotropy. Indeed, the magneto-transport properties below on a 125 nm thick sample confirm this prediction.

**Magnetic, magneto-transport and electro-transport properties of FG2T:** The magnetic property measurements of the new FG2T phase were carried out on polycrystalline sample. **Figure 2 (top right)** shows the temperature dependence magnetic susceptibility measurement in the temperature range 2 K - 300 K and at an applied magnetic field of 0.01 T. The abrupt increase of susceptibility at ca. 250 K ($T_C$) indicates a sharp ordering transition, as observed for strong ferromagnets, and hints at homogeneous composition throughout the sample (see XRD analysis above). Also, the above reported FGT impurity in this sample is found in this measurement, as an inflection is seen at ca. 220 K ($T_C$ of FGT). Consequently, the higher $T_C$ value is due to the new FG2T phase, as further confirmed by transport measurements on single crystal flakes below. The M-H curve (**Figure S5**) recorded at 4 K shows a hysteresis with soft ferromagnet behavior for the FG2T polycrystalline sample like FGT, and both materials become hard ferromagnets in thin flake single crystals (see below).

The resistivity of a 125 nm FG2T single crystal flake, plotted in **Figure 2 (bottom right),** shows a slight decrease from 405 µΩ.cm at 300 K to 400 µΩ.cm at 252 K then a more rapid decrease to 350 µΩ.cm at 2 K. The steep drop at 252 K coincides with the ferromagnetic phase transition. The FG2T resistivity corresponds to metallic behavior as predicted above by DFT. Similar resistivity values at 300 K have been reported for FGT (648 µΩ.cm [11] and 440 µΩ.cm [9b]) and Fe$_{5-x}$GeTe$_2$ (300 µΩ.cm [14b]). FG2T is therefore a ferromagnetic metal as predicted by DFT.

The recorded hysteresis loops of the anomalous Hall resistivity $\rho_H$ for a single crystal FG2T device with thickness of 114 nm are displayed in **Figure 3 (middle and right)** for different temperatures ranging from 2 K to 260 K (device image is shown in the inset **Figure S6**). The $\rho_H$ loops are squared from 2 K up to 160 K with increasing coercive field $H_c$ as the temperature is decreased. $H_c$ reaches ca. 2.6 kOe at 2 K, indicating strong PMA, as predicted by DFT. Above 160 K, the $\rho_H$ loops start to collapse, and disappear at ca. 252 K ($T_C$). Simultaneously, the magnitude of $\rho_H$ loops, i.e., the height between the two saturated values, decreases as the temperature is raised, and vanishes at the Curie temperature $T_C$ (252 K) as illustrated in **Figure S6**. Another evaluation of $T_c$ was performed using the Arrott plot, on the 114 nm device, which gave $T_C$ = 253.8 K (**Figure S7**). The overall temperature dependence of $\rho_H$ of Fe$_{5-x}$Ge$_2$Te$_2$ resembles its mean-field magnetization (see **Figure S6**), but it is slightly steeper.

In summary, we have discovered a new van der Waals (vdW) material, Fe$_{5-x}$Ge$_2$Te$_2$ (FG2T), which was studied experimentally and theoretically. FG2T has a 30 K higher $T_C$ than FGT and shows large perpendicular magnetic anisotropy (PMA). Monolayer FG2T would be an interesting experimental target as it is expected to have a higher $T_C$ value than FGT (130 K). Furthermore, we expect FG2T to enable fundamental physics through heterostructure and near room temperature spintronic device fabrications. During the review of this work, we became aware of the phase $Fe_{2.3}GeTe$ (CCSD 1953048), which has the same space group as FG2T and was reported as a conference proceeding abstract. [25]

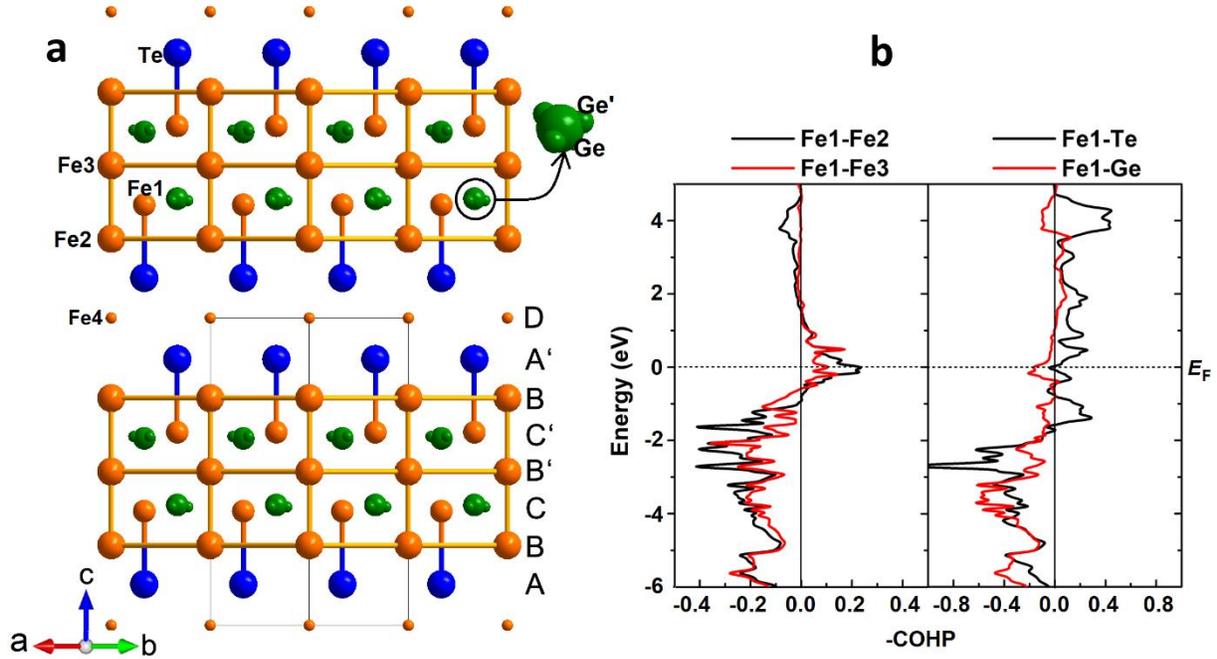

**Figure 1. a**: Projection along [110] of the crystal structure of $Fe_{5-x}Ge_2Te_2$. **b:** LDA -COHP curves for Fe-based interactions in $Fe_{5-x}Ge_2Te_2$. The relative sizes of the spheres represent the site occupation factor: 100% for Te, Fe2 and Fe3; 83.8% Fe1; 16.4% Fe4; 73% Ge and 8.9% Ge'.

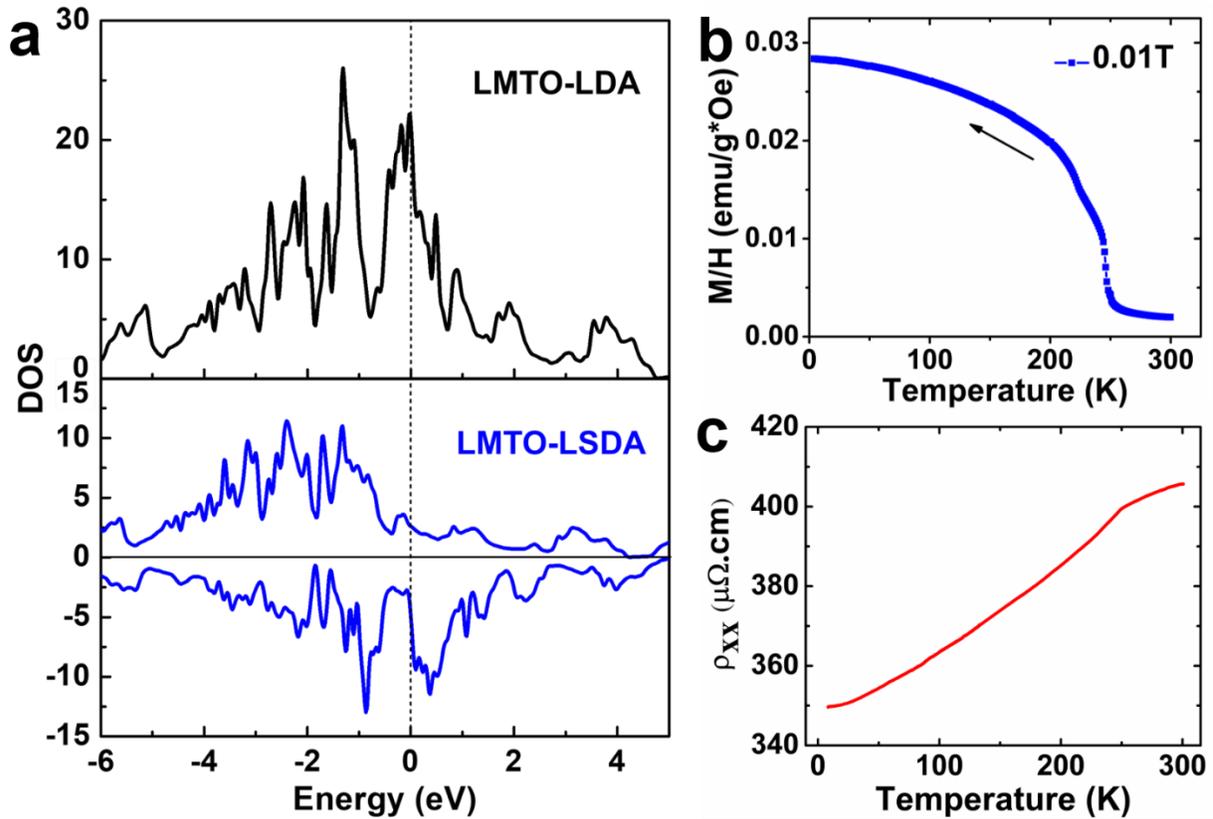

**Figure 2. a:** LDA and LSDA (from LMTO code) density of states (DOS) for $Fe_5Ge_2Te_2$. **b:** Temperature dependence of the magnetic susceptibility of a bulk $Fe_{5-x}Ge_2Te_2$ sample (containing a $Fe_{3-x}GeTe_2$ impurity, small inflection at ca. 220 K). **c:** Temperature dependence of the longitudinal resistivity $\rho_{xx}$ of a 125 nm thick $Fe_{5-x}Ge_2Te_2$ single-crystal device.

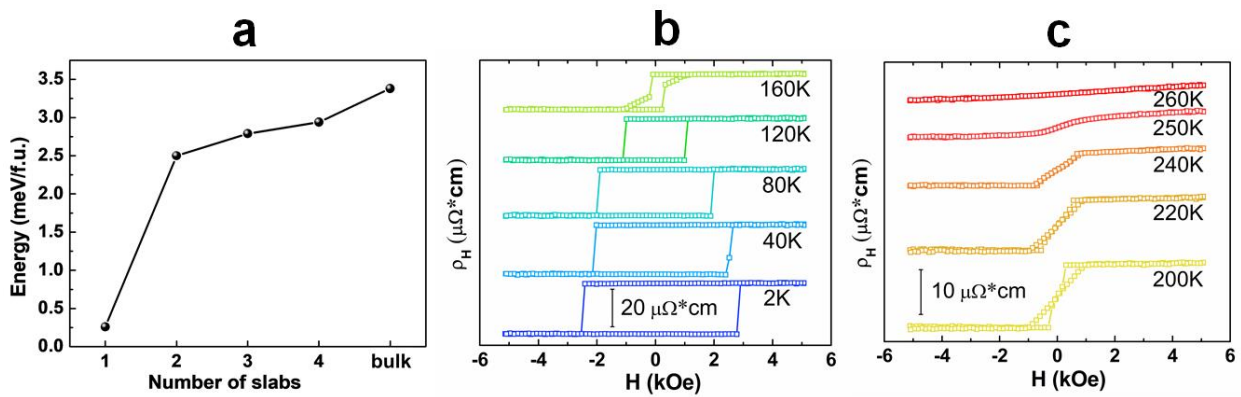

**Figure 3. a**: DFT-calculated magnetic anisotropy energy (MAE, absolute values) of $Fe_5Ge_2Te_2$ plotted as function of number of slabs. **b, c:** Hall resistivity as a function of applied field for a 114 nm thick single-crystal flake of $Fe_{5-x}Ge_2Te_2$ at temperatures from 2 K to 260 K.

**Acknowledgements**

This work was supported by the startup fund to BPTF at UC Riverside and the National Science Foundation Career award to BPTF (no. DMR-1654780). The authors further acknowledge support by DOE BES Award No. DE-FG02-07ER46351 for device nanofabrication and transport measurements. The UC Riverside's High-Performance Computing Center (HPCC) and the San Diego Supercomputing Center (XSEDE) are acknowledged for theoretical calculations.